 \documentclass[final,authoryear,5p]{elsarticle}

\usepackage{epsfig}

\usepackage{amssymb}

\usepackage[a4paper=true,%
breaklinks=true,%
colorlinks=true,%
pdfauthor={A.\ ud-Doula \& Y.\ Naz\'e},%
pdftitle={X-ray emission from MCWs}%
]{hyperref}

\journal{Advances in Space Research}

\begin{document}

\newcommand{\aapr}{A \& A Rv}
\newcommand{\acta}{Act. Astron.}
\newcommand{\araa}{Ann. Rev. Astron. Astrophys.}
\newcommand{\aj}{AJ}
\newcommand{\apj}{ApJ}
\newcommand{\apjs}{ApJS}
\newcommand{\apjl}{ApJ}
\newcommand{\mnras}{MNRAS}
\newcommand{\aap}{A\&A}
\newcommand{\aaps}{A\&AS}
\newcommand{\apss}{Ap\&SS}
\newcommand{\pasp}{PASP}
\newcommand{\pasj}{PASJ}
\newcommand{\jgr}{J. Geophys. Res.}
\newcommand{\planss}{Plan. Space Sci.}
\newcommand{\solphys}{Sol. Phys.}
\newcommand{\nat}{Nature}
\def\<<{{\ll}}
\def\>>{{\gg}}
\def\wig{{\sim}}
\def\spose#1{\hbox to 0pt{#1\hss}}
\def\ltwig{\mathrel{\spose{\lower 3pt\hbox{$\mathchar"218$}}
     R_{\rm A}ise 2.0pt\hbox{$\mathchar"13C$}}}
\def\gtwig{\mathrel{\spose{\lower 3pt\hbox{$\mathchar"218$}}
     R_{\rm A}ise 2.0pt\hbox{$\mathchar"13E$}}}
\def\+/-{{\pm}}
\def\=={{\equiv}}
\def\mubar{{\bar \mu}}
\def\mustar{\mu_{\ast}}
\def\Lambar{{\bar \Lambda}}
\def\Rstar{R_{\ast}}
\def\Mstar{M_{\ast}}
\def\Lstar{L_{\ast}}
\def\Tstar{T_{\ast}}
\def\gstar{g_{\ast}}
\def\vth{v_{th}}
\def\grad{g_{rad}}
\def\glines{g_{lines}}
\def\Mdot{\dot M}
\def\mdot{\dot m}
\def\yr{{\rm yr}}
\def\ksec{{\rm ksec}}
\def\kms{{\rm km/s}}
\def\qad{\dot q_{ad}}
\def\qlines{\dot q_{lines}}
\def\solar{\odot}
\def\Msun{M_{\solar}}
\def\msbyr{\Msun/\yr}
\def\Rsun{R_{\solar}}
\def\Lsun{L_{\solar}}
\def\Be{{\rm Be}}
\def\Rpole{R_{p}}
\def\Req{R_{eq}}
\def\Rmin{R_{min}}
\def\Rmax{R_{max}}
\def\Rstag{R_{stag}}
\def\vinf{V_\infty}
\def\Vrot{V_{rot}}
\def\Vcrit{V_{crit}}
\def\half{{1 \over 2}}
\def\ch{{\it Chandra}}
\def\xmm{{\it XMM}}
\newcommand{\lxlb}{$L_{\rm X}/L_{\rm BOL}$}
\newcommand{\loglxlb}{$\log[L_{\rm X}/L_{\rm BOL}]$}
\newcommand{\beq}{\begin{equation}}
\newcommand{\eeq}{\end{equation}}
\newcommand{\beqa}{\begin{eqnarray}}
\newcommand{\eeqa}{\end{eqnarray}}
\def\phip{{\phi'}}

\begin{frontmatter}

\title{Magnetically Confined Wind Shocks  in X-rays \\ - a Review}

\author{Asif ud-Doula\corref{cor}}
\address{Penn State Worthington Scranton, 120 Ridge View Drive, Dunmore, PA 18512, USA}
\cortext[cor]{Corresponding author - theory}
\ead{asif@psu.edu}

\author{Ya\"el Naz\'e\corref{cor2}\fnref{footnote3}}
\address{Institut d'Astrophysique et de G\'eophysique, Universit\'e de Li\`ege, Quartier Agora, All\'ee du 6 Ao\^ut 19c, Bat. B5C, B4000-Li\`ege, Belgium}
\cortext[cor2]{Corresponding author - observations}
\fntext[footnote3]{Research Associate FRS-FNRS}
\ead{naze@astro.ulg.ac.be}

\begin{abstract}

A subset ($\sim$ 10\%) of massive stars present strong, globally ordered (mostly dipolar) magnetic fields. The trapping and channeling of their stellar winds in closed magnetic loops leads to {\it magnetically confined wind shocks} (MCWS), with pre-shock flow speeds that are some fraction of the wind terminal speed. These shocks generate hot plasma, a source of X-rays. In the last decade, several developments took place, notably the determination of the hot plasma properties for a large sample of objects using \xmm\ and \ch, as well as fully self-consistent MHD modelling and the identification of shock retreat effects in weak winds. Despite a few exceptions, the combination of magnetic confinement, shock retreat and rotation effects seems to be able to account for X-ray emission in massive OB stars. Here we review these new observational and theoretical aspects of this X-ray emission and envisage some perspectives for the next generation of X-ray observatories. 

\end{abstract}

\begin{keyword}
X-rays \sep MHD \sep Massive Stars
\end{keyword}

\end{frontmatter}

\parindent=0.5 cm

\section{Introduction}

Hot luminous, massive stars of spectral type O and B are prominent sources of X-rays which can  originate from three distinct sources: shocks in their high-speed radiatively driven stellar winds, wind-wind collisions in binary systems and magnetically confined wind shocks. 

In  single, non-magnetic O stars, the intrinsic instability of wind driving by line-scattering leads to embedded wind shocks that are thought to be the source of  their relatively soft ($\sim$0.5\,keV) X-ray spectrum, with a total X-ray luminosity that scales with stellar bolometric luminosity, $L_{\rm x} \sim 10^{-7} \times L_{\rm bol}$  \citep{Chl1989, Ber1997, Naz2011}. In  massive binary systems the collision of the two stellar winds at up to the wind terminal speeds can lead to even higher $L_{\rm x}$, generally with a significantly harder (up to 10\,keV)  spectrum (for a review, see Rauw \& Naz\'e, this volume).

Here we discuss a third source of X-rays from OB winds, namely those observed from the subset ($\sim$10\%) of massive stars with strong, globally ordered (often significantly dipolar) magnetic fields \citep{pet13};  in this case, the trapping and channeling of the stellar wind in closed magnetic loops leads to {\em magnetically confined wind shocks}  (MCWS) \citep{BabMon1997a,BabMon1997b}, with pre-shock flow speeds that are some fraction of the wind  terminal speed, resulting in intermediate energies for the shocks and associated X-rays  ($\sim$2\,keV). A prototypical example is provided by the magnetic O-type  star $\theta^1$~Ori~C, which shows moderately hard X-ray emission with a rotational phase variation that matches well the expectations of the MCWS paradigm  \citep{Gag2005}.

Here, we discuss theoretical aspects of magnetic confinement that determine the extent of the influence of the field over the wind. We, then, describe an effect called `shock retreat', which can moderate the strength of the X-rays, or even quench it altogether in extremely low mass loss rate stars, and the effects of rotation. In \S 3, we describe current results in the X-ray observations of early-type magnetic stars, while \S 4 briefly discusses future outlooks of MCWS in X-ray astronomy.

\section{Theoretical Perspective}
To explain X-ray emission from the  Ap/Bp star IQ Aur \citet{BabMon1997a} introduced  the MCWS model. In their approach, they effectively prescribed a fixed magnetic field geometry to channel the wind outflow (see also \citealt{ShoBro1990}). For large magnetic loops, wind material from opposite footpoints is accelerated to a substantial fraction of the wind terminal speed (i.e., $\ge$1000 km s$^{-1}$) before the channeling toward the loop tops forces a collision with very strong shocks, thereby heating the gas to temperatures (10$^7$ - 10$^8$ K) that are high enough to emit hard (few keV) X-rays. This star has a quite strong field ($\sim$4 kG) and a rather weak wind, with an estimated mass loss rate of about $\sim 10^{-10} M_\odot$ yr$^{-1}$, and thus  indeed could be reasonably modeled within the framework of prescribed magnetic field geometry. However, the actual X-ray emission from the star, which is predominantly soft, is further influenced by an effect called `shock retreat' that we describe below.  Later, \citet{BabMon1997b} applied this model to explain the periodic variation of X-ray emission of the O7 star $\theta^1$~Ori~C, which has a lower magnetic field ($\sim$1100 G) and significantly stronger wind (mass-loss rate $\sim 10^{-7} M_\odot$ yr$^{-1}$), raising now the possibility that the wind itself could influence the field geometry in a way that is not considered in the simple fixed-field approach.

\subsection{Magnetic Confinement}
In an interplay between magnetic field and stellar wind,
the dominance of the field is determined  by how strong it
is relative to the wind. To understand the
competition between these two, \citet{udDOwo2002} defined a characteristic
parameter for the relative effectiveness of the magnetic fields in
confining and/or channeling the wind outflow. Specifically, consider
the ratio between the energy densities of field vs. flow,
\begin{eqnarray}
\eta (r, \theta) &\equiv& \frac{B^2/8\pi}{\rho v^2/2} 
\approx \frac{B^2 r^2}{\dot{M}v(r)} \label{etadef}
\\
&=& \left [\frac{B_{\ast}^2 (\theta )
{R_{\ast}}^2}{\dot{M}v_{\infty}}\right ] \left
[\frac{(r/R_{\ast})^{-2n}}{1-R_{\ast}/r} \right ] \, , \nonumber
\end{eqnarray}
where the latitudinal variation of the surface field has the dipole
form given by $B_\ast^2 (\theta) = B_o^2 ( \cos^2 \theta + \sin^2
\theta/4 )$.
In general, a magnetically channeled outflow will have a complex
flow geometry, but for convenience, the second equality in eqn.~(\ref{etadef}) simply characterizes the wind  strength in terms of a
spherically symmetric mass loss rate $\dot{M}=4\pi r^2 \rho v$. The
third equality likewise characterizes the radial variation of
outflow velocity in terms of the phenomenological velocity law $v(r)
=v_{\infty}(1-R_{\ast}/r)^\beta$, with $v_{\infty}$ the wind terminal
speed and $\beta=1$; this equation furthermore models the magnetic field strength
decline as a power-law in radius, $B(r)
=B_{\ast}(R_{\ast}/r)^{(n+1)}$, where, e.g., for a  dipole
$n=2$.

With the spatial variations of this energy ratio thus isolated
within the right square bracket, we see that the left square bracket
represents a dimensionless constant that characterizes the overall
relative strength of field vs. wind. Evaluating this in the region
of the magnetic equator ($\theta=90^o$), where the tendency toward a
radial wind outflow is in  most direct competition with the tendency
for a horizontal orientation of the field, one can thus define an
equatorial `wind magnetic confinement parameter',
\begin{eqnarray}
\eta_{\ast} &\equiv&
\frac{B_\ast^2 (90^\circ) {R_{\ast}}^2} {\dot{M}v_{\infty}}
= 0.4  \,  \frac{B_{100}^2 \, R_{12}^2}{\dot{M}_{-6} \, v_8}.
\label{wmcpdef}
\end{eqnarray}
where $\dot{M}_{-6} \equiv \dot{M}/(10^{-6}\, M_{\odot}$/yr),
$B_{100} \equiv B_o/(100$~G), $R_{12} \equiv R_{\ast}/(10^{12}$~cm),
and $v_{8} \equiv v_{\infty}/(10^8$~cm/s). 
 In order to have any confinement, $\eta_\ast \ge 1$.
As these stellar and wind
parameters are scaled to typical values for an OB supergiant, e.g.
$\zeta$ Pup, the last equality in eqn. (\ref{wmcpdef}) immediately
suggests that for such winds, significant magnetic confinement or
channeling should require fields of order  few hundred G. By
contrast, in the case of the sun, the much weaker mass loss (${\dot
M}_\odot \sim 10^{-14}~M_{\odot}$/yr) means that even a much weaker
global field ($B_{o} \sim 1$~G) is sufficient to yield $\eta_{\ast}
\simeq 40$, implying a substantial magnetic confinement of the solar
coronal expansion. But in Bp stars the magnetic field strength can be of order kG with 
${\dot M}_\odot \sim 10^{-10}~M_{\odot}$/yr leading $\eta_{\ast} \le 10^6$. 
Thus, the confinement in Bp stars is very extreme.

It should be emphasized that  $\dot{M}$ used in the above 
formalism is a value obtained for a spherically symmetric non-magnetic wind  as the magnetic field may significantly influence the predicted circumstellar density and velocity structure. 

\subsubsection{Alf\'ven Radius}
The extent of the effectiveness of magnetic  confinement
is set by the Alfv\'en radius, $R_A$, where flow and Alfv\'en velocities are equal. This will also determine
the extent of the largest loops and thus the highest shock velocities affecting the hardness of X-ray emission.
This radius can be derived from eqn.~(\ref{etadef}) where the second square bracket factor
shows the overall radial variation;
$n$ is the power-law exponent for radial decline of
the assumed stellar field, e.g. $n=2$ for a pure dipole,
and  with $v(r)= v_\infty (1-R_\ast/r)^\beta$ where $\beta$ is the velocity-law index (typically $\beta \approx 1$).
For a star with a non-zero field, we have $\eta_{\ast} > 0$, and so given
the vanishing of the flow speed at the atmospheric wind base, this
energy ratio always starts as a large number near the stellar surface,
$\eta(r \rightarrow R_{\ast}) \rightarrow \infty$.
But from there outward it declines quite steeply, asymptotically as
$r^{-4}$ for a dipole, crossing unity at the
Alfv\'{e}n radius defined implicitly by  $\eta(R_A) \equiv 1$.

For a canonical $\beta=1$ wind velocity law,
explicit  solution for $R_{\rm{A}}$ along the magnetic equator
requires finding the appropriate root of
\begin{equation}
    \left ( {R_A \over R_\ast } \right )^{2n} -
    \left ( {R_A \over R_\ast } \right )^{2n-1} = \eta_{\ast}
\, ,
\label{radef}
\end{equation}
which for integer $2n$ is just a simple polynomial, specifically
a quadratic, cubic, or quartic for $n =$~1, 1.5, or 2.
Even for non-integer values of $2n$, the relevant solutions
can be approximated (via numerical fitting)
to within a few percent by the simple general expression,
\begin{equation}
\frac{R_{\rm{A}}}{R_{\ast}}
\approx 1  + (\eta_{\ast} + 1/4)^{1/(2n)} -  (1/4)^{1/(2n)}
\, .
\label{raapp}
\end{equation}
For weak confinement, $\eta_{\ast} \ll 1$, we find
$R_{\rm{A}} \rightarrow R_{\ast}$,
while for strong confinement, $\eta_{\ast} \gg 1$, we obtain
$R_{\rm{A}} \rightarrow \eta_{\ast}^{1/(2n)} R_{\ast}$.
In particular, for the standard dipole case with $n=2$,
we expect the strong-confinement scaling
$R_{\rm{A}}/R_{\ast} \approx \eta_{\ast}^{1/4}$.

Clearly $R_{\rm{A}}$ represents the radius at which the wind speed $v$
exceeds the local Alfv\'{e}n speed $V_{A}$.
It also characterizes the maximum radius where the
magnetic field still dominates over the wind. For  Ap/Bp stars where stellar fields are of order kG,
$\eta_\ast \gg 1$, e.g. for $\sigma$~Ori~E it is about $10^7$,
implying an Alfv\`{e}n radius $\sim 60 R_\ast$.  Thus, in Bp (and Ap) stars wind is trapped
to large radii creating extensive magnetospheres. This also implies that X-rays from Bp stars should be intrinsically hard.  But as
we show in \S 2.2, shock retreat effects may soften it significantly.

\subsubsection{MHD Simulations}

The initial magnetohydrodynamic (MHD) simulations by \citet{udDOwo2002} assumed, for
simplicity, that radiative heating and cooling would keep the wind
outflow nearly isothermal at roughly the stellar effective temperature. 
The simulations studied the dynamical competition
between field and wind by evolving  MHD simulations
from an initial condition when a dipole magnetic field is suddenly introduced into a previously relaxed,
one-dimensional spherically symmetric wind.

\begin{figure*}
\begin{center}
\vfill
\includegraphics[scale=0.15]{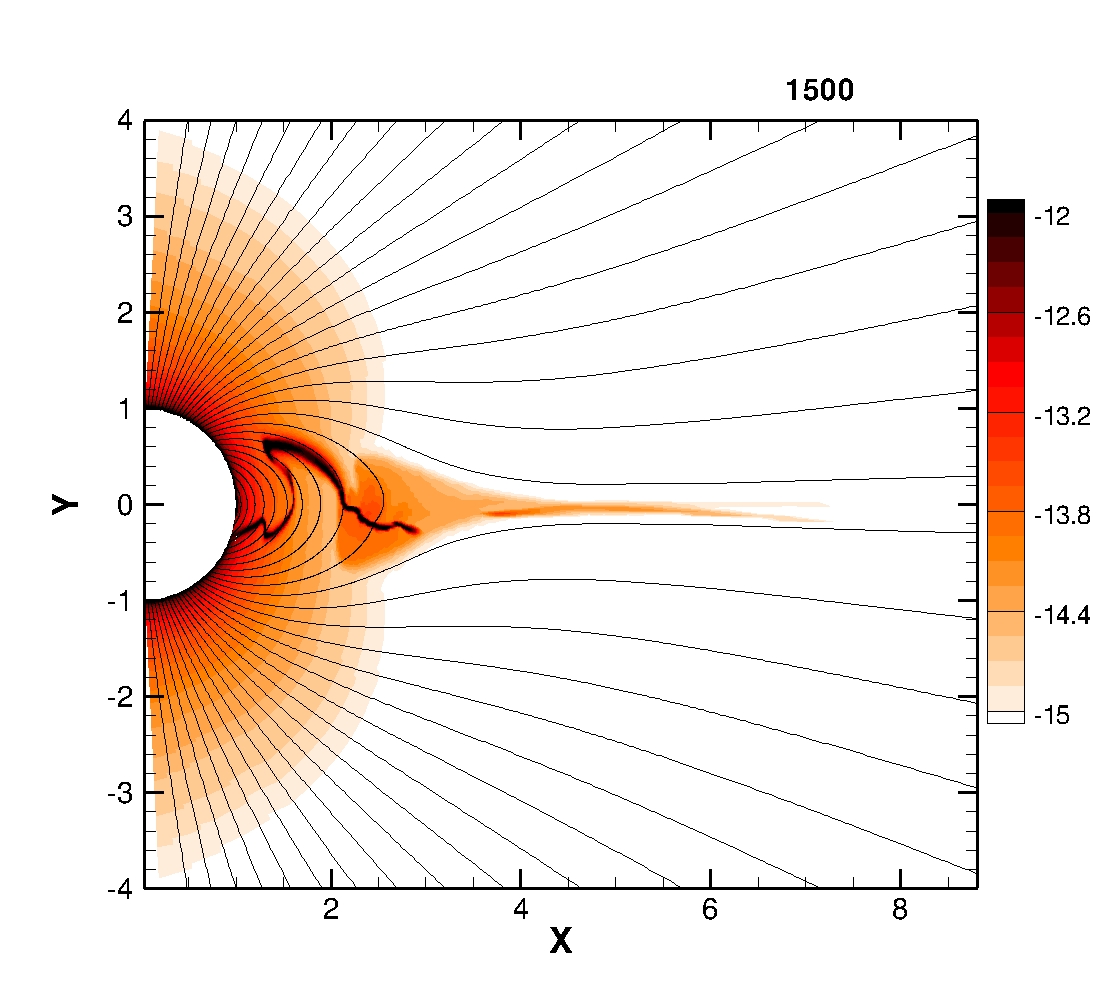}
\includegraphics[scale=0.15]{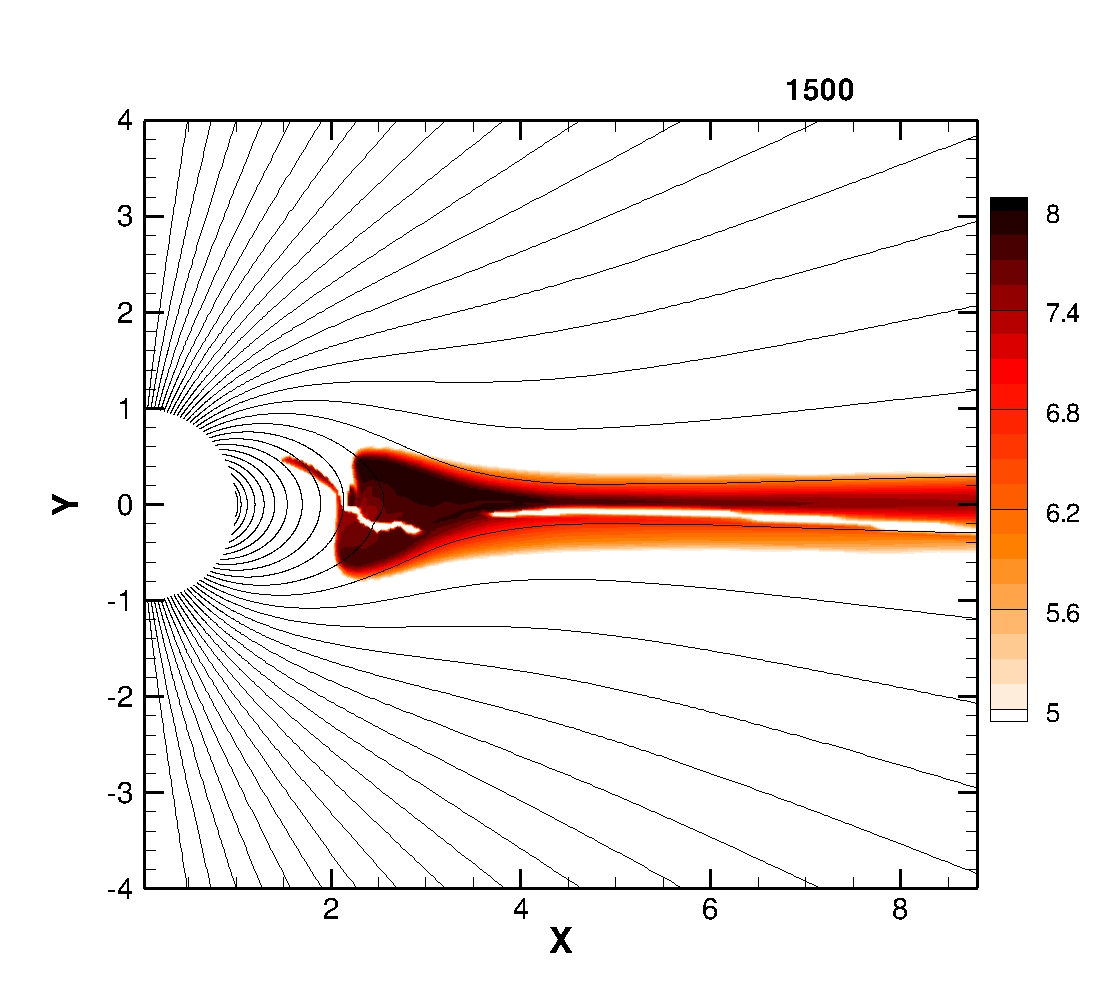}
\includegraphics[scale=0.15]{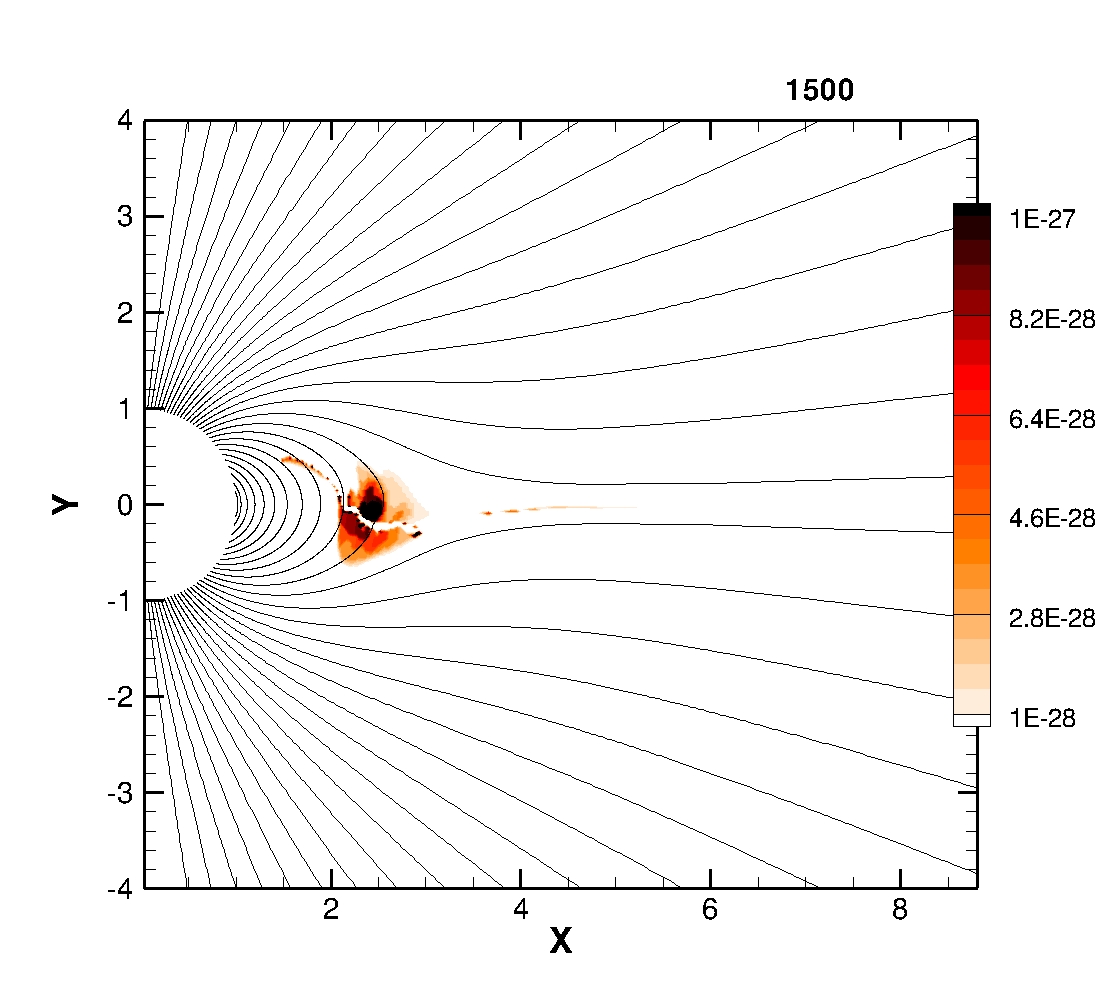}
\caption{
Colour plots of log density (left) and log temperature (middle) for arbitrary snapshot of structure in the standard model with $\eta_\ast =100$.
The right panel plots the proxy X-ray emission $XEM_{T_x}$  (weighted by the radius $r$),  on a {\em linear} scale for a threshold X-ray temperature $T_x = 1.5$\,MK. Figure reproduced from  \citet{udd14}.
}
\label{fig:rhotxem}
\end{center}
\end{figure*}

Immediately after the introduction of the field, the dynamic interplay between the wind and the field leads to 
two distinct regions. Along the polar region, the wind freely streams radially outward, stretching the field lines into a radial configuration, as can 
be inferred from the left panel of illustrative Fig. \ref{fig:rhotxem}. If the field is strong enough, around the magnetic equator a region of closed magnetic
loops is formed wherein the flow from opposite hemispheres collides to make strong shocks, quite similar to what was predicted in the semi-analytic, fixed-field models of 
\citet{BabMon1997a}.  The shocked material forms a dense disk-like structure which is opaque to line-driving. 
But, its support against gravity by the magnetic tension along the convex field lines is inherently unstable, leading to a complex pattern of fall back along the loop lines down to the star,  again as suggested by the left panel of  Fig.  \ref{fig:rhotxem} .

Note that even for weak field models with moderately small confinement, $\eta_{\ast} \le 1/10$,  the field still has a noticeable global influence
on the wind, enhancing the density and decreasing the flow speed near
the magnetic equator. However, shock speeds are probably not sufficiently large enough to produce any X-rays in that case.

But to model the actual X-ray emission from shocks that form from  the magnetic
channeling  and confinement, subsequent efforts \citep[][]{udD2003, Gag2005}
have relaxed  the assumption of isothermal equation of state in earlier studies to include a detailed energy equation
that follows the radiative cooling of shock-heated material.
The magnetically channeled wind shock model provides excellent agreement with the diagnostics from the phase-resolved {\it Chandra} spectroscopy of $\theta^1$~Ori~C (see \S 3 for more details).

\subsubsection{3D MHD}
\begin{figure}
\includegraphics [scale=0.32]{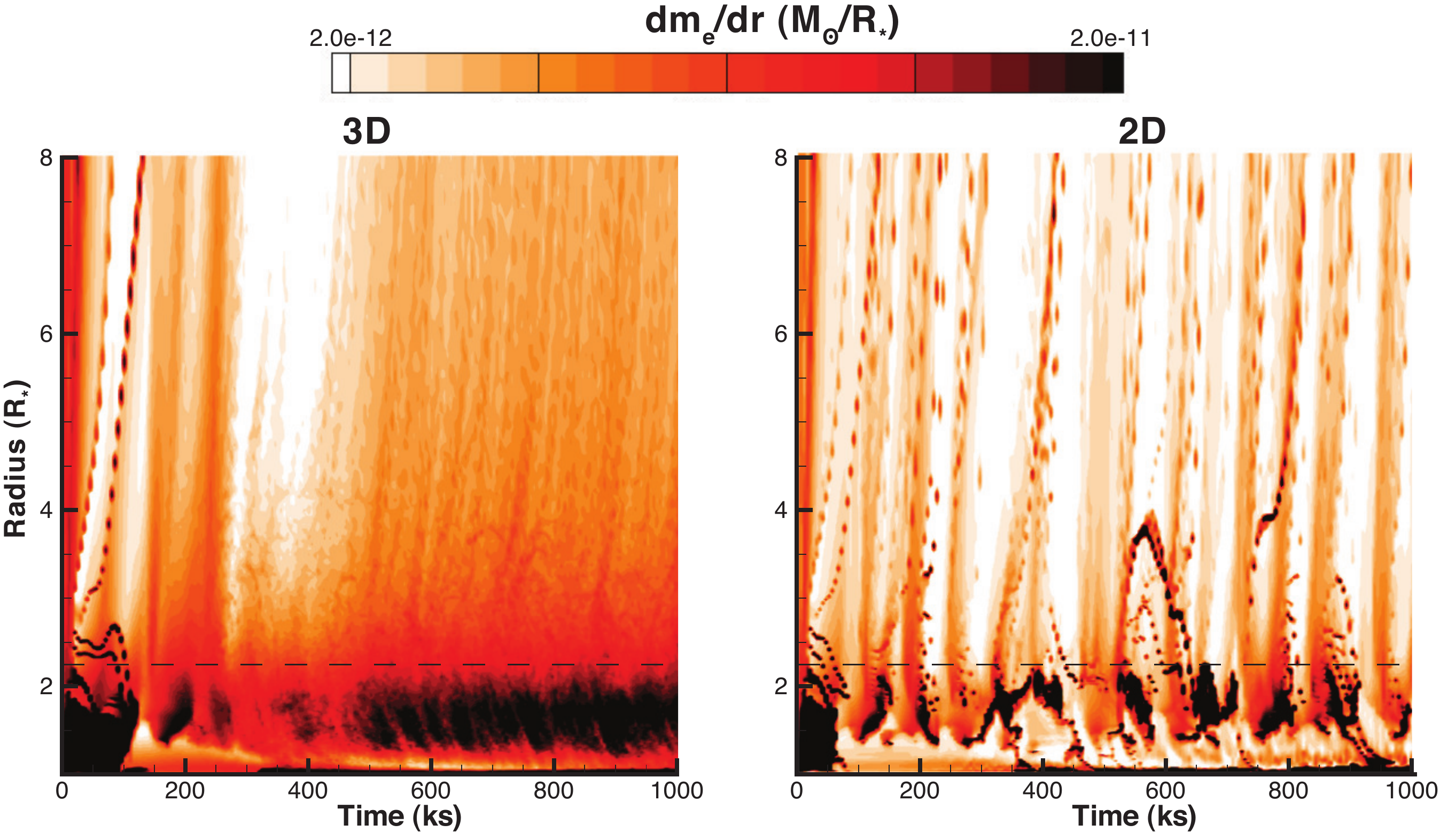}
\caption
{Equatorial mass distribution $dm_{\rm e}/dr$ (in units $M_\odot/R_\ast$) for the azimuthally averaged 3D model (left) and for a corresponding 2D model (right), plotted versus radius and time, with the dashed horizontal line showing the Alfv\`en radius $R_{\rm A} \approx 2.23 R_*$.
Following a similar adjustment to the initial condition,  the long-term evolution of the 2D models is characterized by a complex  pattern of repeated strings of infall, whereas the 3D azimuthally averaged model settles into a relatively smooth asymptotic state characterized by an enhanced mass near and below the Alfv\'{e}n radius.
Figure reproduced from  \citet{udD2013}.}
\label{dmdr-r-vs-t}
\end{figure}

To get a more complete picture of the wind-field interaction, one needs ultimately full 3D MHD simulations. Using the specific parameters  chosen to represent the prototypical slowly rotating magnetic O star $\theta^1$~Ori~C, 
for which centrifugal and other dynamical effects of rotation are negligible, \citet{udD2013} have computed the first fully 3D MHD model of its wind. The computed global structure in latitude and radius resembles that found in previous 2D simulations, with unimpeded outflow along open field lines near the magnetic poles, and a complex equatorial belt of inner wind trapping by closed loops near the stellar surface, giving way to outflow above the Alfv\'en radius. In contrast to this previous 2D work, the 3D simulation  also shows how this complex structure fragments in azimuth, forming distinct clumps of closed loop infall within the Alfv\'en radius, transitioning in the outer wind to radial spokes of enhanced density with characteristic azimuthal separation of 15$^o-20^o$. Applying these results in a 3D code for line radiative transfer, they show that emission from the associated 3D `dynamical magnetosphere'  (DM) matches well the observed H$\alpha$ emission seen from $\theta^1$~Ori~C, fitting both its dynamic spectrum over rotational phase and the observed level of cycle-to-cycle stochastic variation. Comparison with previously developed 2D models for the Balmer emission from a dynamical magnetosphere generally confirms that time averaging over 2D snapshots can be a good proxy for the spatial averaging over 3D azimuthal wind structure, as illustrated in Fig. \ref{dmdr-r-vs-t}. Nevertheless, fully 3D simulations will still be needed to model the emission from magnetospheres with non-dipole field components such as $\tau$ Sco. A similar study examining X-rays  predicted by this 3D MHD model is currently under way (Cohen et al., in preparation).

\subsection {Shock Retreat}

\begin{figure}
\begin{center}
\includegraphics[scale=0.5]{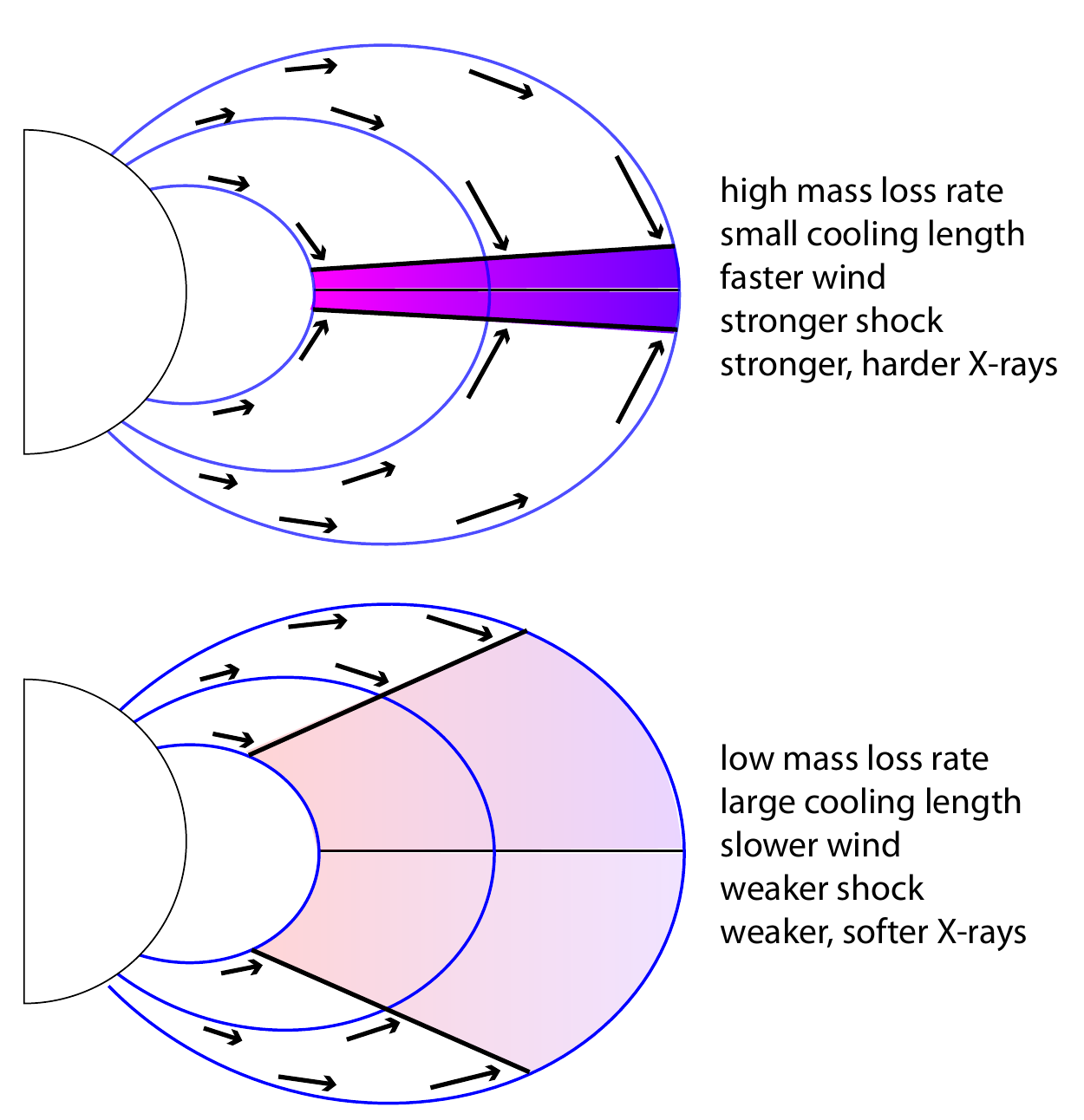}
\caption
{Schematic illustration of the ``shock retreat'' from inefficient cooling associated with a lower mass loss rate ${\dot M}$, showing a hemispheric, planar slice of a stellar dipole magnetic field.
Wind outflow driven from opposite foot-points of closed magnetic loops is channeled into a collision near the loop top, forming magnetically confined wind shocks (MCWS).
For the high ${\dot M}$ case in the upper panel, the efficient cooling keeps the shock-heated gas within a narrow cooling layer, allowing the pre-shock wind to accelerate to a high speed and so produce strong shocks with strong, relatively hard X-ray emission.
For the low ${\dot M}$ case in the lower panel, the inefficient cooling forces a shock retreat down to lower radii with slower pre-shock wind,  leading to weaker shocks with weaker, softer X-ray emission. Figure reproduced from  \citet{udd14}.}
\label{fig:shockretreat}
\end{center}
\end{figure}
Although magnetic confinement is the dominant effect in production of X-rays in magnetic massive stars, there are other secondary effects that can influence both the total luminosity and hardness of X-rays.  In particular, an effect called `shock retreat' directly related to cooling efficiency can significantly moderate X-rays in stars with low mass loss rates such as Bp stars.

As illustrated schematically in Figure \ref{fig:shockretreat}  \citep[see also Figure 13 of \citealt{BabMon1997a}]{udd14}, the much lower mass loss rates of such B-stars also implies much less efficient cooling of the post-shock flow. When the associated cooling length, becomes comparable to the Alfv\'{e}n radius, the shock location is effectively forced to ``retreat'' back down the loop, to a lower radius where the lower wind speed yields a weaker shock, implying then a much softer X-ray spectrum.  It was shown that such shock retreat can be effectively parametrized by a cooling parameter, $\chi_\infty \propto V_\infty^4/\dot{M}$.  A high value of  this cooling parameter, $\chi_\infty$, implies an inefficient cooling, and, thus, a large shock retreat effect  \citep{udd14}.

 In extreme low mass loss cases, value of $\chi_\infty \gg 1$ which may lead the shock retreat to be as large as the loop size. This can quench the wind  within the closed magnetosphere suppressing the X-ray production altogether.  Here, it may  be interesting to note that confined winds were also suggested to play a role in the X-ray emission of mid-B to mid-A stars, when detected \citep{cez06,ste11,rob11}. Indeed, after examining the ratio of X-ray to radio luminosities, \citet{rob11} excluded low-mass companions as the sources of X-rays. Considering a sample of Ap/Bp stars, these authors (see also \citealt{rob14}) further found that only stars brighter than $\sim200L_{\odot}$ emit X-rays, implying the presence of a threshold in physical phenomena, whose origin could be linked to shock retreat effects.

Now, to estimate the total X-ray luminosity in typical OB stars, $L_X$,   \citet{udd14} developed a semi-analytic analysis to derive a generalized ``X-ray Analytic Dynamical Magnetosphere'' (XADM) scaling law for X-rays emitted from such MCWS. To describe the paradigm briefly without much technical details, XADM assumes local latitudinal variation of radial mass flux at the stellar surface as prescribed by \cite{OwoudD2004}. The size of the magnetosphere can be readily estimated from the magnetic confinement parameter, $\eta_*$. It then computes shock velocity along a loop. Energy emitted as X-rays above a threshold can be estimated from kinetic energy dissipated in the shocks. Integration over the closed magnetosphere then yields the total X-ray luminosity above a threshold assuming 100\% efficiency.

The resultant XADM scaling follows a very similar trend to the full magnetohydrodynamic (MHD) simulation results, but is about a factor 5 higher. Compared to the idealized steady-state emission of the analytic XADM model, the numerical MHD simulations show extensive time variability with repeated intervals of infall of cooled, trapped material, and it appears that this lowers the overall efficiency of X-ray emission to about 20 per cent of the idealized XADM prediction. 

\subsection{The Effects of Rotation}

A further complication arises from  rotation (not accounted for by XADM), as short-period modulations and substantial broadening in photospheric spectral lines \citep{ConEbb1977,Fuk1982}, corresponding to projected rotation speeds of hundreds of km/s, have been detected in some massive stars.

For a non-magnetic rotating star, conservation of angular momentum in
a wind outflow causes the azimuthal speed near the equator to decline
outward as $v_{\phi} \sim 1/r$, meaning that rotation effects tend to be of
diminishing importance in the outer wind. However, at the stellar surface, reduced effective gravity leads to enhanced mass loss rate \citep{udD2008,udD2009} and slower wind.

By contrast, in a rotating star with a sufficiently strong magnetic field, magnetic torques on the wind can spin it up; for some region near the star, i.e.,
up to about the maximum loop closure radius $R_{\rm{c}}$ (closely related to
Alfv\`en radius, $R_A$), they can even maintain a nearly rigid-body rotation. As such, even for a star with surface rotation below the orbital
speed, maintaining rigid rotation will eventually lead to a balance between the outward centrifugal force from rotation and the inward force of gravity.

Unsupported material at radii $r < R_{\rm{K}}$, Kepler radius,  will tend to fall back toward the
star, but any material maintained in rigid-rotation to radii
$r > R_{\rm{K}}$ will have a centrifugal force that {\em exceeds} gravity, and
so will tend to be propelled further outward, unless held down by strong enough magnetic field as is the case for Bp and some other  magnetic B stars.

This complex interaction between the field and the wind naturally leads to two distinct groups of magnetic stars:  ``dynamical magnetospheres'' (DM) and ``centrifugal magnetospheres'' (CM) \citep{Sun2012, pet13}. For the former case, $R_A< R_{\rm{K}}$  and rotation is dynamically unimportant with all the supported material within the magnetosphere falling back toward the star. Notably, all Of?p stars fall within this category  (their slow rotation probably comes from magnetic braking, see more details below). X-ray emission from DM is governed mainly by the magnetic confinement and shock retreat effects. 

Whereas, for the latter case, $R_A>R_{\rm{K}}$ leading to co-rotating magnetospheres, as is the case for the rapidly rotating strong magnetic star  $\sigma$~Ori~E. The effect of rapid rotation on X-ray emission is currently a very much work in progress. However, preliminary studies show that the enhanced mass loss due to rotation leads to a more efficient cooling, which in turn can both increase and harden X-rays \citep{Bar2015}. Here, there is a three way competition between rotation, magnetic confinement and shock retreat. Since the cooling parameter $\chi_\infty \propto V_{shock}^4/\dot{M}$, the shock retreat effects might be  moderated by the effects of rotation which increases $\dot{M}$ in general. But which effect ultimately dominates may depend on the details of stellar parameters of specific stars  and rapidly rotating B stars can have  a variety of X-ray luminosities. This is discussed further in \S 3.

Another natural consequence of interaction of magnetic field with wind and rotation
is the increase in angular momentum loss from the stellar surface.
It turns out that the angular momentum is carried away by not only the gas,
but also by the field itself in the form of Maxwellian stress tensor.
In a classical analysis, \citet{WebDav1967} showed that the total angular momentum
loss  can be expressed as ${\dot J} = (2/3) {\dot M} \Omega R_{A}^{2}$,
where ${\dot M}$ is the mass loss rate, $\Omega$ is the stellar angular
velocity, and $R_{A}$ is a characteristic Alfv\'{e}n radius. \citet{udD2009} have shown that the typical timescale for spindown for massive magnetic O stars is about 1 Myr or so. This has a direct effect on  the evolution of these stars and  their  X-ray luminosities. Currently, all known strongly magnetic O-stars are slow rotators, with the exception of Plaskett's star where interactions within the binary are speculated to play an additional role in spinup of the primary \citep{Gru2013,PalRau2014}.

\section{Observational Perspective}

From the theoretical perspective,  magnetic confinement, shock retreat and rotation effects are the main ingredients for production and explanation of X-rays in MCWS. Of course, other factors like the tilt angle of the field to the rotation axis or more generally the field geometry can also have substantial influence, as could also the presence of a companion, of stellar pulsations or of additional (azimuthal) structures in the wind.  Now, let us turn our attention to what the observations actually show. We shall see that, despite a few exceptions, indeed these three parameters can mostly account for X-ray emissions from magnetic massive stars.

The observed scatter in X-ray luminosity of AB stars may, as in OB stars,  be related to different confinement, rotation, and cooling parameter values, but better knowledge of these parameters are required before a comprehensive comparison with models can be done just like for OB stars, so we will not describe these objects in detail here.

\subsection{X-ray luminosity}

The X-ray luminosity of massive OB stars is well known to scale linearly with the bolometric luminosity, \loglxlb $\sim-7$. However, there are some exceptions that display overluminosities, like the well-known cases of $\theta^1$\,Ori\,C or $\tau$\,Sco \citep{sch00,Gag2005,mew03,coh03}. The discovery of magnetic fields in these objects is relatively recent, but they were found to play a key role in explaining their X-ray spectra. This overluminosity, of about one dex compared to the majority (``normal'') of OB stars, was further found in many other magnetic massive stars ($\sigma$\,Ori\,E - \citealt{ski08} or the Of?p stars - \citealt{naz04,naz07,naz08,naz15, Pet2015}). However, some other magnetic massive stars appear much fainter \citep{fav09,osk11,ign13,naz14}. 

Several attempts have been made recently to understand such diversity. A first, limited survey of 11 objects by \citet{osk11} found no link between the level of X-ray luminosity and the magnetic field strength, rotational period, bolometric luminosity, or pulsational period (when existing). The general trend amongst stars (more/less luminous) seemed however qualitatively accounted for by the formula of \cite{BabMon1997b}, $L_{\rm X}\propto \dot M V_\infty B^{0.4}$. 

\begin{figure*}
\begin{center}
\vfill
\includegraphics[scale=0.3]{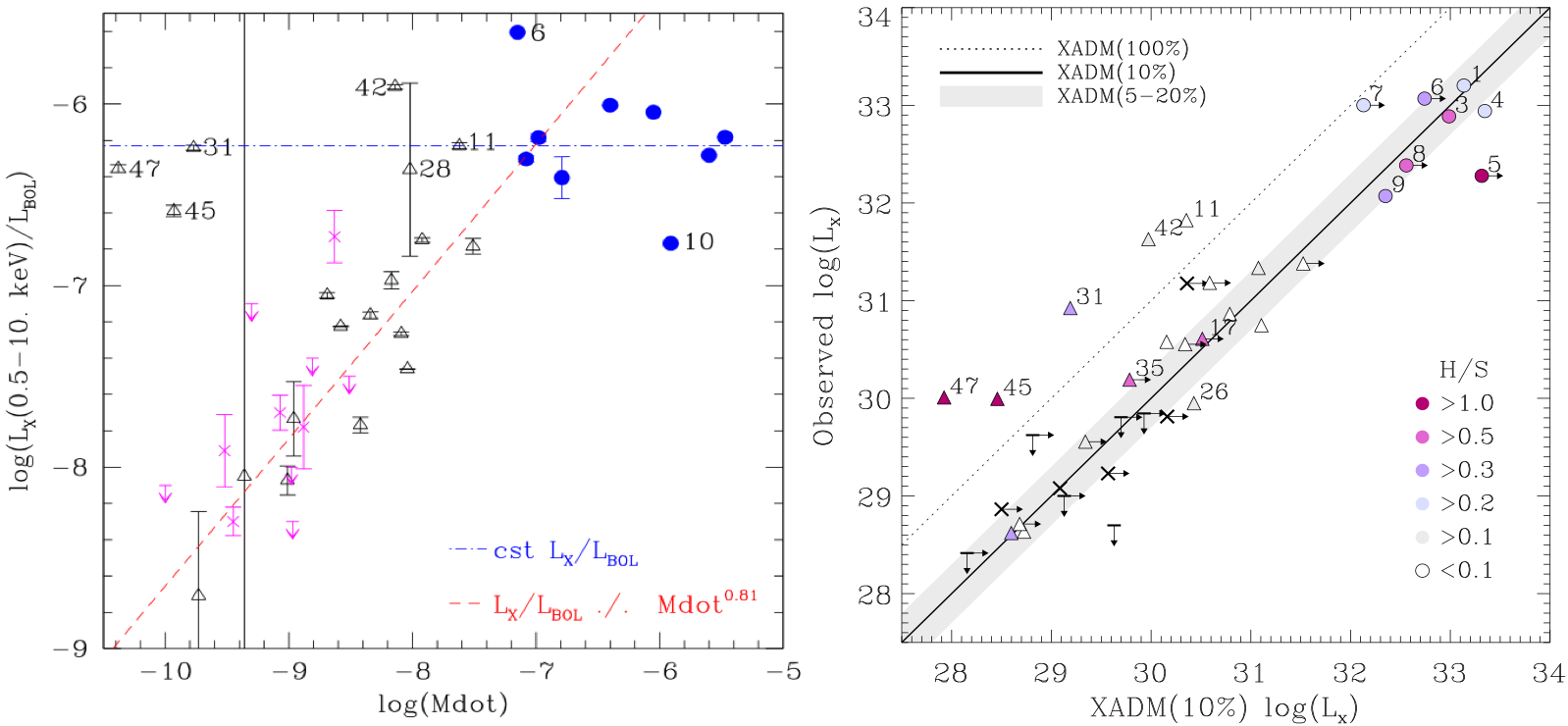}
\caption
{Observed X-ray luminosity (corrected for ISM absorption) as a function of mass-loss rate (left) and of the predicted values using the XADM model of \citet{udd14} (right). {\it Left:} Filled blue dots correspond to O-stars, black empty triangles to B-stars, magenta crosses and downward-pointing arrows to faint detections and upper limits on the X-ray luminosity, respectively. Blue and magenta lines represent $L_x \propto \dot{M}^{0.6}$ and $L_x \propto \dot{M}^{1.4}$ relation. {\it Right:} The dotted line illustrates the ideal model with 100\% efficiency whereas the solid line indicates a scaling by 10\%; the grey shaded area corresponds to scalings by 5-20\% (a range in efficiency consistent with MHD models). The symbol shapes are the same as in the left panel, but this time symbols are colour-coded according to their measured hardness ratios. Figures reproduced from \citet{naz14}, numbers correspond to star IDs defined in \citet{pet13}. }
\label{fig:Lx-Obs}
\end{center}
\end{figure*}

A larger survey, covering nearly two thirds of the catalog of known magnetic OB stars of \citet{pet13} was undertaken by \citet{naz14}. The 40 targets used in the survey were scattered all over the so called ``magnetic confinement-rotation diagram''  \citep[see][]{pet13}, limiting the biases. The situation then appeared in more detail: some of the sources (O-stars and a few B-stars) have \loglxlb $\sim-6.2$ or equivalently  $L_{\rm X}\propto \dot M^{0.6}$ whereas other B-stars display $L_{\rm X}\propto L_{\rm Bol}^{1.9}$ or $L_{\rm X}\propto \dot M^{1.4}$ (Fig. \ref{fig:Lx-Obs}). 

These observations are clearly at odds with the earlier Babel \& Montmerle prediction of $L_{\rm X}\propto \dot M$. Moreover, the level of emission predicted by this formalism is 1.8\,dex too high.  However, the more recent XADM formalism, confirmed by MHD simulations, has  shown much more promise \citep{udd14}. The right panel of figure~\ref{fig:Lx-Obs} shows a comparison between the observed X-ray luminosity of magnetic stars  and the predicted values using the XADM model of \citet{udd14}. The dotted line illustrates the ideal model with 100\% efficiency whereas the solid line indicates a scaling by 10\%; the grey shaded area corresponds to scalings by 5-20\%, a range in efficiency consistent with MHD models. The observed X-ray luminosities for most of the known magnetic stars fall within this range.

However, there are several clear exceptions (Fig. \ref{fig:Lx-Obs}), five of which show observed luminosities larger than predicted. A first outlier is $\tau$ Sco (\#11) which has a complex non-dipolar field, something the current version of XADM does not consider as it is limited to pure dipolar cases. Three cases, $\sigma$~Ori~E (\#31), HD\,182180 (\#45) and HD\,142184 (\#47), are rapid rotators \citep{pet13}. However, XADM does  not  take into account rapid rotation, but \citet{Bar2015} have shown that rotation both enhances and hardens the X-ray emission. It may be noted in this context that those three objects are also particularly hard \citep{naz14}, but it should also be mentioned that the difference between rotating and non-rotating cases in Bard \& Townsend models is less than one dex, while the observed difference for those three objects is larger (about 2 dex). In addition, HD\,64740, a star with similar properties (rotation rate, confinement, cooling) as $\sigma$~Ori~E does not display a bright and hard emission, and its $L_{\rm X}$ appears in line with XADM predictions. Finally, a last exception, HD200775 (\#42) remains unexplained.

There is also one case where the observed luminosity is smaller than the predicted one. It corresponds to NGC 1624-2 (\#5), which hosts a very strong (20\,kG) magnetic field. It turns out that the huge magnetosphere of this extreme magnetic star heavily attenuates its X-ray emission, making it appear less luminous and much harder than it intrinsically is \citep{Pet2015}. In fact, the estimated intrinsic luminosity is quite consistent with prediction from the XADM formalism.

\subsection{Absorption and hardness of spectrum}

Traditionally, a basic ``magnetic-thus-hard'' paradigm was assumed. However, one could a priori consider observed hardness of MCWS with care. Indeed, for NGC 1624-2, the apparent hardness of  the X-ray spectrum is actually due to the large absorption by the dense circumstellar material ($N_{\rm H}\sim1-4\times10^{21}$\,cm$^{-2}$), the intrinsic hardness being drastically different \citep{Pet2015}. MHD simulations of such an extreme magnetosphere  are presently out-of-reach, but the general properties of this object qualitatively agree with those expected from confined winds in a extreme magnetosphere as predicted by XADM paradigm \citep{Pet2015}. 

However, NGC 1624-2 clearly is an extreme case. In the global survey of magnetic OB stars by \citet{naz14}, the amount of local absorption in spectral fits was compared to that found from general surveys of OB stars and they were found to be similar. This suggests no significantly increased absorption due to confined winds in most magnetic OB stars, which is corroborated by the monitoring of $\theta^1$\,Ori\,C, where only a small increase (from 4.5 to $5\times10^{21}$\,cm$^{-2}$, not even formally significant considering the errors) was found \citep{Gag2005}. This is consistent with the small density columns predicted by MHD models \citep[see e.g. Fig.~10 of ][]{Pet2015}. The observed hardness, once corrected for ISM absorption, is thus generally a good indicator of the intrinsic hardness. 

Observationally, the spectrum of $\theta^1$\,Ori\,C (\#3 in Fig.~\ref{fig:Lx-Obs}) is dominated by plasma at $\sim$2.5\,keV \citep{sch00,Gag2005}, and hot plasma is certainly present in other stars, e.g. $\tau$\,Sco \citep{mew03,coh03} or Of?p stars \citep{naz04,naz07,naz08}. But the hard component does not always dominate, as in Of?p stars (which usually appear soft), or may not even be detectable \citep[e.g. $\beta$\,Cep, ][]{fav09}. Even  cases of ``twins'' wherein a pair of stars have twin-like similar stellar parameters, revealed some surprises, for instance with one $\tau$\,Sco analog being quite similar to its prototype (HD\,63425) while the other one (HD\,66665) was softer \citep{ign13}. 

Surveys by \citet[11 stars,][]{osk11} and \citet[40 stars,][]{naz14} have noted such diversity  of the situation. Comparison with general OB stars surveys further reveals that magnetic O-stars appear at least somewhat harder than ``normal'' O-stars while magnetic B-stars  seemed rather softer than usual B-stars \citep{naz14}. The latter conclusion may however be biased by the different population sampled, the observed magnetic B-stars sampling lower X-ray luminosities than in surveys. Nevertheless, no correlation could be found between hardness and magnetic or stellar parameters \citep{naz14}. In particular, the prediction of harder X-ray emission in case of stronger magnetic confinement, larger mass-loss rate, or higher rotation rate was not detected.

\subsection{X-ray Variability}
 
By design, theoretical predictions such as XADM are time independent, although fully self-consistent MHD models do suggest certain level of time variability in X-rays and other bands. Observationally, several types of variabilities associated with magnetically confined winds have been detected. In general, these variabilities do not change global properties of X-ray emission discussed above, but they do show how complex the interplay between the field and wind can be. 
%

\subsubsection{Flares}
The first type of possible variations is short-term X-ray flaring. MHD models of dynamical magnetospheres show a complex temporal behaviour, with stochastic episodes of heating and infall back on the star - but they are limited to small regions, hence are smoothed out in 3D magnetospheres \citep{udd13}. In centrifugal magnetospheres, however, the trapped material accumulates and, to keep the balance, some ejection needs to take place, through ``centrifugal'' breakouts involving magnetic reconnection on the outer edges of magnetospheres \citep{tow05,udD2006}. This should generate flares and such events have been recorded in $\sigma$\,Ori\,E by ROSAT, \xmm, and \ch\ \citep{GroSch2004,San2004,ski08,cab09}, but their interpretation was much debated: is it intrinsic to $\sigma$\,Ori\,E or due to an unseen low-mass companion? Several arguments can be put forward for the former scenario. For example, the decay part of the flare lightcurve suggests a size of $\sim1R_*$ for the region involved in the event (considering $R_*$ to be the radius of $\sigma$\,Ori\,E), which is reasonable in the context of MCWS \citep{GroSch2004,mul09} but this size would correspond to 10--20$R_*$ when considering the radius of a K-type companion, too large for a flare associated to an active coronal region in the companion \citep{mul09}. In addition, while flares are typical for coronal sources and their luminosity is compatible with that observed, flares of the same magnitude as those observed occurred rarely for low-mass stars and their typical recurrence timescale would have prevented several observation of such events over two decades \citep{mul09}. On the other hand, \citet{bou09} did detect the presence of a low-mass companion and the analysis of the source position during the \ch-HRC flaring episode favors the low-mass companion as origin of the flare \citep{pet12}. Furthermore, the magnetospheric mass derived by \citet{tow13} appears to be two orders of magnitude below the level required for centrifugal breakouts to happen as initially suggested by \citet{udD2006}. A rather similar situation was found for the magnetic Ap star IQ\,Aur, where the estimated size of the flaring region, the position of the X-ray source, the soft character of the recorded X-rays  (compared to young active stars), and the high X-ray to radio luminosity ratio favor an origin of the flare in the Ap star, but the influence of a low-mass companion cannot be formally excluded \citep{rob11}. Finally, the detection of several flares in HD\,47777 was also reported by \citet{naz14} but again, it remains to be confirmed that they are actually associated with the magnetically confined winds of the star, rather than with an yet unknown low-mass companion of the star. 

\subsubsection{Rotational modulation}
A second type of variation is rotational modulation, expected to occur on longer timescales than flaring. This type of variability strongly depends on the magnetospheric geometry and the orientation of the star with respect to the line of sight. Observations have been analyzed to reveal such changes.

Amongst magnetic OB stars, $\tau$\,Sco displays a rather complex field geometry, with one model restricted regions of densely concentrated closed loops asymmetrically distributed on the stellar surface \citep{Don2006a}. Strong phase-locked variability could thus be expected \citep[40\% modulation, ][]{Don2006a}, as some regions appear in plain view at some phases and are occulted at others, but a dedicated Suzaku monitoring found at best very low amplitude changes in flux \citep[a few percents and mostly in soft band, ][]{ign10}. This may be explained if the distribution of hot gas is different (e.g. in smaller loops) from the magnetospheric geometry derived from optical spectropolarimetry. Certainly more detailed investigation is needed before this observational result can fully be understood.

With a simpler, mostly dipolar magnetic field, other magnetic OB stars should show changes related to the viewing angle on the confined winds (i.e. alternative edge-on/face-on view). Indeed, X-ray variations simultaneous with optical changes have been detected in at least four well-documented cases: $\theta^1$\,Ori\,C \citep[and references therein]{Gag2005,ste05}, HD\,191612 \citep{naz10}, CPD\,$-28^{\circ}$2561 \citep{naz15}, and NGC1624-2 \citep{Pet2015}. For these objects, the X-ray emission and the H$\alpha$ emission appear brighter when the confined winds are seen face-on. 

 In the X-ray domain, variation amplitudes $(F_{max}/F_{min}-1)$ typically reach 40--100\% for $\theta^1$\,Ori\,C , HD\,191612, NGC1624-2, and CPD\,$-28^{\circ}$2561 as well as NU\,Ori \citep{naz14} and Tr16-22 \citep{naz14tr1622}, but the amplitude appears smaller in a few other cases. For example, outside flares, the  X-ray lightcurve of $\sigma$\,Ori\,E shows limited variability: \citet[ see in particular their Fig. 5]{ski08} detected variations with an amplitude $(F_{max}/F_{min}-1)\sim30$\% but at the 1.4\% significance level only - the amplitude is in fact comparable to the noise and requires confirmation. Similarly, the X-ray flux of $\beta$\,Cep changes by at most 10\% and its phase-locked nature still need to be established \citep{fav09,naz14}. Finally, a stable X-ray emission is observed for HD\,148937, which is not surprising: it is always seen nearly pole-on, hence the view on the confined winds of this oblique rotator does not vary much, explaining the low amplitude changes in the optical domain and the absence of detectable variations in X-rays \citep{naz14}. 

The phase-locked variability  observed in X-rays could be explained in two different ways. On the one hand, it could be related to absorption. For example, if X-rays arise in the dense equatorial regions, then some absorption of that emission could take place, reducing the observed (soft) X-ray flux especially when these regions are seen edge-on. However, to produce a large flux decrease, this absorption increase would become noticeable in spectral fitting. This is the case of the extremely magnetic O-star NGC1624-2, where the very dense magnetosphere produces a strong and variable additional absorption ($1-4\times10^{22}$\,cm$^{-2}$, \citealt{Pet2015}). The spectrum with minimum emission clearly lacks soft X-ray photons when compared to that of the maximum emission phase, and occultation effects (see below) could only be responsible at most for a 20\% variation in (intrinsic) flux. However, this situation is exceptional, as the local, additional absorption due to confined winds generally remains low, as mentioned before, and no significant increase in absorption for the edge-on configuration was reported in other stars \citep{Gag2005,naz14}. 

On the other hand, flux variations in magnetic stars could be related to regular occultation of the emission regions by the stellar body. This is the favored explanation and it could explain the scatter in amplitude change. For example, the small variations of $\beta$\,Cep are compatible with occultation of hot plasma located at 4--6$R_*$, the position derived from high-resolution spectra \citep{fav09}. $\sigma$\,Ori\,E has a very large magnetosphere (about 31R$_*$ for the Alfv\'{e}n radius), which minimizes the impact of occultations and could contribute to explain the lack of clear variation detection \citep{naz14}. However, as the flux changes are often accompanied by changes in spectral shape, this would imply that occultation affects plasma of different temperatures in different ways as could be expected, e.g., if thermal stratification exists \citep{naz14}. In this context, it may be worth noting the variety of behaviours, which remains to be explained: HD\,191612, Tr16-22, CPD\,$-28^{\circ}$2561, and NU\,Ori appear harder when brighter, while $\theta^1$\,Ori\,C appears to be softer while brighter \citep{naz14}. 

\begin{figure}
\begin{center}
\includegraphics[width=9.2cm]{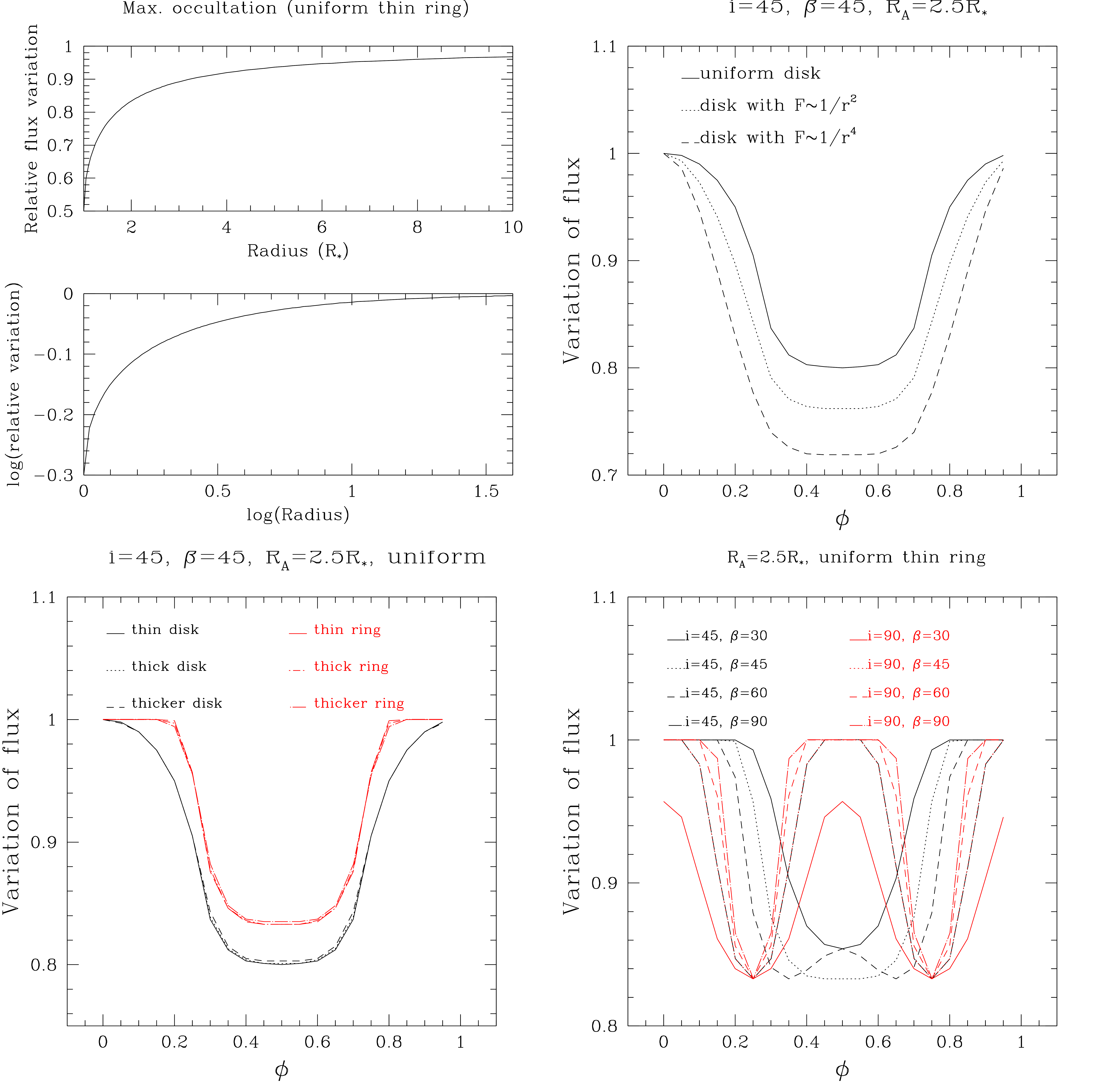}
\caption{ {\it Top left:} Evolution with radius of the minimum value of the apparent flux, relative to the unocculted case, of a ring of negligible $\Delta r$ lying in the magnetic equatorial plane. Note that, depending on orientation, structures very close to the star may remain always partially occulted throughout the cycle, hence the y-axis value does not necessarily represent the observed $F_{\rm min}/F_{\rm max}$, it is only the minimum possible. {\it Top right:} Variation of relative flux with phase, for different brightness functions (uniform brightness, evolution as $r^{-2}$ or $r^{-4}$). The Alfv\'{e}n radius, inclination, and obliquity are typical of $\theta^1$\,Ori\,C. {\it Bottom left:} Variation of relative flux with phase, for different geometries. ``Disk'' is a structure extending from $R_*$ to $R_A$, ``ring'' an annuli extending from $R_A-R_*$ to $R_A$; ``Thin'' corresponds to structures lying in the magnetic equator only, ``thick'' applies to structures with a total thickness of 0.3\,$R_*$, ``thicker'' for thicknesses of 0.5\,$R_*$. The Alfv\'{e}n radius, inclination, and obliquity are typical of $\theta^1$\,Ori\,C. {\it Bottom right:} Variation of relative flux with phase, for different combinations of inclination and obliquity but a similar geometry (thin ring, with $R_A=2.5R_*$). This agrees well with observations, e.g. the single minimum of $\theta^1$\,Ori\,C  vs. the two minima of CPD\,$-28^{\circ}$2561.}
\label{figvar}
\end{center}
\end{figure}

To assess the occultation explanation, we can model the confined winds by a simple optically-thin structure near the star, and predict the occultation degree for various situations. We tested several geometries: a disk up to Alfv\'{e}n radius $R_A$ and a ring from $R_A-R_*$ to $R_A$, which better reproduce the geometry of emitting regions observed in MHD simulations. Figure \ref{figvar} shows how occultation depends on geometry (a disk is more occulted than a ring) and its variation with phase depends on orientation (i.e. for various inclination and obliquity combinations). Most important is the location of the plasma, of course. Considering a ring geometry in agreement with MHD models yields a relative flux variation $F_{\rm min}/F_{\rm max}$ of $\sim0.83$, or $[F_{\rm max}/F_{\rm min}]-1\sim 20$\%, for the case of $\theta^1$\,Ori\,C while the observed value is $[F_{\rm max}/F_{\rm min}]-1\sim50$\%. Similar conclusions are reached for CPD\,$-28^{\circ}$2561 or HD\,191912. To get higher occultation degrees, one could change the plasma location, as structures closer to the star are more occulted than those further away. To get the observed value of $\theta^1$\,Ori\,C, one would however need to consider a ring of negligible radius located at 1.15$R_*$,  less than half the Alfv\'{e}n radius but also much less than the plasma location derived by He-like triplets \citep{Gag2005b}, which seems dubious. For cases like Tr16-22, where $[F_{\rm max}/F_{\rm min}]-1\sim 100$\%, then the hot plasma would need to be at the photosphere. Clearly, the simple model is underevaluating occultation, most probably because of its simplicity. Considering some limited absorption may help, but a fully optically-thick case would probably be inappropriate because of its phase shift\footnote{\citet{Don2001} have envisaged the case of confined winds being fully optically thick. In that case, the X-ray flux would be maximum when the confined winds are seen edge-on (i.e. when the optical emission are minimum) as the emission from both hemispheres would then be seen, while only the emission of ``upper half'' of the confined winds would be observable at other phases, that of the ``lower half'' being fully absorbed. While flux variations by a factor of two have been detected, it must be noted that a shift by half a cycle between X-ray and optical maxima has not been reported, up to now.}.

\subsection{Structure of confined winds, as revealed by high-resolution spectra}

High-resolution spectra can yield a wealth of information. With current instrumentation, line widths and shifts can be evaluated with precisions down to a few tens km\,s$^{-1}$ in the most favorable cases (a few hundreds km\,s$^{-1}$ more typically). Furthermore, the comparison of lines from H-like and He-like ions and of components of $fir$ triplets of He-like ions constrain the temperature and location of the emitting plasma. However, such measurements are currently only possible for the brightest X-ray sources, so that few magnetic massive stars have been investigated in this respect ($\tau$\,Sco - \citealt{mew03,coh03}, $\theta^1$\,Ori\,C - \citealt{sch00,Gag2005,Gag2005b}, HD\,191612 - \citealt{naz07}, HD\,148937 - \citealt{naz08,naz12}, $\beta$\,Cep - \citealt{fav09}, IQ\,Aur - \citealt{rob11}). 

Within noise limitations, the X-ray lines of magnetic massive stars were found to be symmetric, and globally unshifted. This agrees well with MHD models. In the case of $\theta^1$\,Ori\,C, global fitting however suggests small variations in velocity \citep{Gag2005}: from $-75$\,km\,s$^{-1}$ when the star is seen pole-on to about 100\,km\,s$^{-1}$ when seen edge-on. This change needs to be confirmed as the errors are large but also because one cannot exclude a stochastic variation when only a single observation per phase is available. If further observations provide evidence that velocity varies with phase, then refinement of models will be needed, as no such changes are currently predicted \citep{Gag2005}.

Reported widths of X-ray lines largely depend on object and ion considered. The narrowest widths, so far, were found for $\beta$\,Cep, whose lines are dominated by instrumental resolution, yielding only upper limit on intrinsic widths \citep[$<$600\,km\,s$^{-1}$, ][]{fav09}. Larger widths, $FWHM\sim 600-800$\,km\,s$^{-1}$, were reported for ions with high ionization potential (Mg, Si, S) in $\tau$\,Sco, $\theta^1$\,Ori\,C, and HD\,148937, three stars with more rapid winds than $\beta$\,Cep. Such widths are much smaller than observed for ``normal'' O-type stars ($FWHM \sim v_{\infty}$), where lines arise in embedded wind shocks distributed all over the wind, hence cover a larger velocity range. They indicate formation in slowly-moving plasma, in agreement with the confined winds scenario. However, most MHD models predict even narrower lines \citep{Gag2005}. 

Furthermore, lines from ions with lower ionization potential, notably oxygen, appear broader \citep[$FWHM\sim1800-2000$\,km\,s$^{-1}$][]{Gag2005,naz07,naz08}. These lines are associated with cooler plasma, which could have a different origin than hotter plasma. For example, the dominant hot plasma in $\theta^1$\,Ori\,C is thought to arise in confined winds while the cooler one could arise in embedded wind shocks as in normal O-stars \citep{Gag2005}. This dual origin could be supported by the different temperatures derived from the different ions \citep{sch00}. In Of?p stars, however,  the spectra are dominated by the cooler component, i.e. confined winds emit soft X-rays in these objects (see above), but it cannot be excluded that current errors, which are large, are somewhat blurring the picture.

In He-like triplets, the forbidden line is suppressed when the density is high or the UV radiation is intense. In the case of massive stars, the latter effect is the most important one and, thanks to dilution with distance, enables us to locate the emitting region. In $\tau$\,Sco, $\theta^1$\,Ori\,C, HD\,148937, and $\beta$\,Cep, the start of the emitting region is found to be close to the photosphere,  at radii $r \sim 1.5-3\,R_*$ for the first three stars and $r \sim 4-6\,R_*$ for the latter case. These values are only slightly lower than the corresponding Alfv\'{e}n radii of these stars, they thus appear qualitatively compatible with MHD simulations. In IQ\,Aur, the forbidden line was found to be normal, suggesting a formation radius larger than 7$R_*$ \citep{rob11} - despite a large value, this also appears compatible with the supposed location of confined winds in this star.

\section{Discussion, Conclusion and Future Outlook}

The discovery of magnetism in massive O stars is a relatively recent phenomenon although magnetism in B stars was known for much longer. This led to a number of new observational campaigns such as MiMeS \citep{wad12} or BOB \citep{mor15} that increased the total number of known magnetic OB stars significantly. There have also been some theoretical developments in an attempt to understand this increasingly larger group of stars. In this brief review, we discussed specifically the X-ray emission from these objects.

Although there are some outliers, semi-analytic scaling of XADM paradigm can correctly estimate X-ray luminosities from most magnetic OB stars. A phenomenon called shock retreat can reduce the X-ray luminosity, and for the case of  B stars with extremely low mass-loss rate, it may even quench the wind within the closed magnetosphere. Rotation can counter some of the effects, but much work remains to be done in order to fully understand these complex objects.

As the number of magnetic detections increases with time, future X-ray missions will certainly enlarge the samples. Indeed, most of the results presented here come from a relatively small size sample while a large number of parameters are involved in the confined wind phenomenon and  its observations (magnetic strength, magnetic geometry, inclination of the system, and stellar parameters such as temperature, rotation rate, wind velocity and mass-loss rate). Notably, only about 10 magnetic O-stars are currently known and have been studied in X-rays. Moreover, the sensitivity of current facilities does not allow a large study of magnetic massive stars at high spectral resolution, nor the detection of very faint X-ray emission (e.g. Ap/Bp stars). This certainly poses a challenge to understand these object fully and with confidence.

The advent of the new X-ray mission Athena+ \citep{nan13} may notably enable us to study confined winds in radically different environments like low-metallicity galaxies such as the Magellanic Clouds, where massive stars are known to display weaker stellar winds. Such new  data would allow to test our current models on vastly different cases, thus improving their reliability. The very first extragalactic magnetic massive candidates have already been proposed \citep{mai01,pau11,naz15mc,wal15}, so that several targets will become available in the next decade for such studies.

Another important observational information comes from monitoring the X-ray emission. It is a natural way to probe the magnetosphere, and it provides strong constraints on its hot plasma content. Unfortunately, such programs are difficult to obtain because of their length and the need of time-constraints, therefore only few objects could be followed in detail. Moreover, when a monitoring exists, it generally covers a single cycle or less - it is thus currently difficult to separate potential stochastic changes from periodic ones. With a better sensitivity, such as foreseen for Athena+, shorter exposures will be needed, reducing the total duration hence facilitating the gathering of multiple exposures.

Finally, high spectral resolution data also provide crucial information by allowing detailed studies of the X-rays associated to confined winds. Currently, only a few such spectra were obtained, and they did not yet allow analyses of the line profiles, as lines are barely resolved. Future facilities such as the European Space Agency's  Athena+ observatory  \citep{nan13} should provide access not only to high spectral resolution for more objects but also to better spectral resolution than currently available. This will allow us to pinpoint the properties of a larger sample with better quality, further constraining the hot plasma distribution and kinematics. Of utmost interest in this context are monitoring performed at high resolution. The behaviour of different lines will then be compared, e.g. those from ions with low-ionization potential and those associated with high ionization potential. Any difference between them will indicate different emission processes, which will be identified thanks to detailed line profile studies. Furthermore, if velocity shifts such as tentatively identified in $\theta^1$\,Ori\,C are confirmed, then detailed tomographic analyses will become possible, enabling the reconstruction of the magnetospheric activity kinematics.

Along with the observational developments, theory must also keep up the pace. Currently, most of the fully consistent MHD simulations are limited to 2D. One exception is the 3D MHD study of H$\alpha$ emission from $\theta^1$~Ori~C by \citet{udD2013}. Since rotation in this case was dynamically unimportant, rotation and field axes were assumed to be aligned. However, in many other cases rotation is dynamically important and the field is tilted with respect to the rotation axis. Such oblique rotator cases must be modelled using full 3D MHD, something currently missing. Moreover, for stars for which magnetic confinement parameter $\eta_\ast \gg 1$, MHD cannot be used due to high Alfv\`en speed that forces Curant time to be impractically small, rendering such numerical simulations essentially impossible. 

In such extreme cases, alternate methods must be employed.  \cite{TowOwo2005}  present  a  semi-analytic method
for modelling the circumstellar environment of early-type stars, which is essentially a generalized extension of Babel and Montmerle
model. They assume that the magnetic field lines remain rigid and corotate with the star without being influenced by the dynamics of the wind. This is essentially equivalent to the assumption of $\eta_\ast \rightarrow \infty$.  This Rigidly Rotating Magnetosphere (RRM) model can be applied to an arbitrary field geometry and tilt angle between the field and rotation axes.  However, there are no dynamic forces involved in this model, instead effective gravitational + centrifugal potential is calculated based on the constrained motion of the plasma.  Accumulation surfaces are assumed to lie along the location of minimum effective potential.

As a further improvement to the RRM model,  \cite{Tow2007} introduced a new Rigid-Field Hydrodynamics (RFHD) method to modelling the circumstellar
environments  of strongly magnetic massive stars as defined above. Just like in the RRM method, the field lines are treated as  rigid, but now the flow along the lines is computed self-consistently using hydrodynamical equations including line force due to radiation.  In this ansatz, flow along each each field line is considered to be an independent 1D flow.  They perform a large number of such 1D calculations for differing field lines, then piece them together to build up a time-dependent 3D model of a magnetosphere.

This method is now superseded by {\it Arbitrary} Rigid-Field Hydrodynamics (ARFHD) method which now allows any configuration of  self-consistent magnetic topology. It also improves some of the  numerical algorithms and radiative cooling and includes thermal conduction \citep{Bar2015}. Since the flow along each field line can be solved independently of other field lines, the computational cost of this approach is a fraction of an equivalent full MHD simulation.  However, this method suppresses any coupling between the flow on adjacent field lines which in practice may have some important ramifications. As such, a new, improved method will be needed in the future.

\section{Acknowledgements}
AuD acknowledges support by NASA through Chandra Award number TM4-15001A and 16200111 issued by the Chandra X-ray Observatory Center which is operated by the Smithsonian Astrophysical Observatory for and behalf of NASA under contract NAS8- 03060. AuD also acknowledges support for Program number HST-GO-13629.008-A provided by NASA through a grant from the Space Telescope Science Institute, which is operated by the Association of Universities for Research in Astronomy, Incorporated, under NASA contract NAS5-26555. YN acknowledge support from  the Fonds National de la Recherche Scientifique (Belgium), the PRODEX XMM contract, and an ARC grant for Concerted Research Action financed by the Federation Wallonia-Brussels.

%

\end{document}